\documentclass[11pt,a4paper]{article}

\usepackage[T1]{fontenc}

\usepackage{jheppub}
\usepackage{mathtools}
 \usepackage{subcaption}

\newcommand{\be}{\begin{eqnarray}}
\newcommand{\ee}{\end{eqnarray}}
\newcommand{\ba}{\begin{array}}
\newcommand{\ea}{\end{array}}

\newcommand{\bea}{\begin{eqnarray}}
\newcommand{\eea}{\end{eqnarray}}

\newcommand{\bi}{\begin{itemize}}
\newcommand{\ei}{\end{itemize}}

\newcommand{\nn}{\nonumber}

\begin{document}

\title{Exact summation of leading logs around $T\bar T$ deformation of $O(N+1)$-symmetric 2D QFTs}

\author[a]{Jonas Linzen,}
\author[a,b]{Maxim V.~Polyakov,}
\author[b,c]{Kirill M. Semenov-Tian-Shansky,}
\author[d,e,f]{\\ Nika S.  Sokolova}


\affiliation[a]{Ruhr University Bochum, Faculty of Physics and Astronomy,
Institute of Theoretical Physics II, D-44780 Bochum, Germany}
\affiliation[b]{National Research Centre ``Kurchatov Institute'': Petersburg Nuclear Physics
Institute, RU-188300 Gatchina, Russia}
\affiliation[c]{Higher School of Economics,
National Research University, RU-194100 St. Petersburg, Russia}		
\affiliation[d]{St.~Petersburg State University, Faculty of Physics, Ul'yanovskaya ul.~3, RU-198504, Peterhof, St.~Petersburg, Russia (until 07.2020)}
\affiliation[e]{Perimeter Institute for Theoretical Physics, Waterloo, ON N2L 2Y5, Canada}
\affiliation[f]{Department of Physics \& Astronomy, University of Waterloo, Waterloo, ON N2L 3G1, Canada}



\abstract{
We consider a
general (beyond $T\bar T$) deformation of the 2D $O(N+1)$ $\sigma$-model by the irrelevant
dimension-four operators. The theory deformed in this most general way is not integrable,
and the $S$-matrix loses its factorization properties.
We perform the all-order summation of the leading infrared logs for the  $2 \to 2$ scattering amplitude and
provide the exact result for the  $2 \to 2$ $S$-matrix in the leading logarithmic approximation.
These results can provide us with new insights into the properties of the theories deformed by irrelevant operators more general than the $T\bar T$ deformation.
}

\maketitle

\section{Introduction}

The study of nonlinear sigma models in two-dimensional spacetime has a long history, dating back to the 70's, when it was discovered
\cite{Polyakov:1975rr,Polyakov:1983tt,Faddeev:1985qu,DAdda:1978vbw}
that this class of theories shares common
non-trivial features with four-dimensional non-abelian gauge theories, while being easier to handle and sometimes even exactly solvable.
For example, for the $O(N+1)$ 2D $\sigma$-model the exact $S$-matrix can be constructed
relying on the integrability of the model, see Ref.~\cite{Zamolodchikov:1978xm} for a review.

Furthermore, it was demonstrated in Ref.~\cite{Smirnov:2016lqw,Zamolodchikov:2004ce}
that the deformation of an integrable theory by the irrelevant composite operator $T\bar T$, built with the components of the energy-momentum tensor,
 can be regarded as a peculiar kind of integrable perturbation
\cite{Cavaglia:2016oda,Conti:2018tca,Conti:2018jho,McGough:2016lol,Cardy:2019qao,Babaei-Aghbolagh:2020kjg}
(see {\it e.g.} Ref.~\cite{Jiang:2019hxb}
for a pedagogical review).

In the present paper we consider the deformation of the integrable $O(N+1)$ (with the field on the sphere $S^N$) 2D $\sigma$-model by the most general operators
of the mass dimension four. The theory deformed in this way loses its integrability property.
For the most general deformation of the theory
scattering amplitudes acquire
leading log (LL) corrections which spoil the factorization property
of the $S$-matrix. In the case of the non-deformed 2D $\sigma$-model the corresponding LL corrections can be absorbed into the running coupling
constant by the renormalization group (RG) methods, see discussion in Chapter~5 of Ref.~\cite{Zamolodchikov:1978xm}. However, the generally deformed theory
is non-renormalizable;
therefore,
the RG methods can not be applied to it directly.

Our goal is to perform the exact summation of the LLs for the $2\to 2$ scattering amplitude in the non-renormalizable $O(N+1)$ 2D $\sigma$-model
deformed by the most general dimension-four composite operators.
The theory we consider is
defined by the following action:
\begin{align}
S= \int d^2 x \left(\frac{1}{2}\partial_\mu \vec{n}\cdot  \partial^\mu \vec{n} -
g_1 (\partial_\mu \vec{n} \cdot \partial^\mu \vec{n})
(\partial_\nu \vec{n} \cdot \partial^\nu \vec{n})-
g_2 (\partial_\mu \vec{n} \cdot \partial_\nu\vec{n})
(\partial^\mu \vec{n} \cdot \partial^\nu \vec{n})\right)
\label{LagrangianBissextile}
\end{align}
with fields $\vec n \in S^N$ of the radius $F$,
{\it i.e.} satisfying the  constraint
$\sum_{A=1}^{N+1} n^An^A=F^2$.
The coupling constants
$g_1$, $g_2$ and $F$ have
the mass-dimensions
\begin{align}
\left[g_1\right]=\left[g_2\right]=-2,\quad\left[F\right]=0.
\end{align}
The action (\ref{LagrangianBissextile}) corresponds to the deformation of the $O(N+1)$ 2D $\sigma$-model by the most general dimension-4 operators.
It can be rewritten in the form:
\begin{align}
S= \int d^2 x \left(\frac{1}{2}\partial_\mu \vec{n}\cdot  \partial^\mu \vec{n} -
4\lambda\ {\rm det}\left(T_{\mu\nu}\right) -\frac{1}{G} \
(\partial_\mu \vec{n} \cdot \partial_\nu\vec{n})
(\partial^\mu \vec{n} \cdot \partial^\nu \vec{n})\right), \label{LagrangianBissextile2}
\end{align}
where $\lambda= g_1$ is the coupling describing the $T\bar T$ perturbation to the $O(N+1)$-symmetric $\sigma$-model. The operator
\begin{align}
T_{\mu\nu}=\partial_\mu \vec{n}\cdot \partial_\nu \vec{n}-\frac 12 \eta_{\mu\nu} (\partial_\alpha \vec{n}\cdot \partial^\alpha \vec{n}),
\end{align}
is the energy-momentum tensor
of the $O(N+1)$ symmetric $\sigma$-model, $\eta_{\mu\nu}={\rm diag(1,-1)}$ is the Minkowski tensor in 2D. The coupling
\begin{align}
\frac1G=2g_1+g_2; \ \ [G]=2
\end{align}
describes a  deviation from  the $T\bar T$ perturbed theory.
The limit $G\to \infty$ corresponds
to the $T\bar T$-deformed $\sigma$-model that was studied {\it e.g.} in Ref.~\cite{Bonelli:2018kik}.

In the limit of zero curvature of the field manifold $S^N$ ($F\to \infty$) the theory (\ref{LagrangianBissextile2}) corresponds to the
$O(N)$-symmetric free field
theory deformed by generic dimension-4 operators  which was recently considered 
in Refs.~\cite{Polyakov:2018rdp,Linzen:2018pvj}\footnote{In these references the corresponding theory
was named as the bissextile or bi-quartic model. The relation to $T\bar T$ deformations was not discussed in these references.}. In \cite{Linzen:2018pvj}
we performed the all-loop exact summation of the leading logs for the $2\to 2$ scattering amplitude in the deformed free field theory. First, it was found that the LLs contribution do not depend on the
coupling constant
$\lambda$ in
(\ref{LagrangianBissextile2}). This reflects the fact that the $T\bar T$-deformed
theory enjoys integrability of the undeformed theory. Second, the explicit non-trivial solutions for the scattering amplitude in the LL-approximation
were obtained. It was shown that the corresponding solutions exhibit non-trivial analytical structure in the variable
$\frac{s}{4 \pi G} \ln(\mu^2/s)$
giving the first example of the non-trivial $S$-matrix beyond the $T\bar T$ deformation.
In this paper we perform the all-order exact summation of LLs for finite radius of the field manifold $S^N$.

It is known that the general deformation of the QFT by irrelevant operators
(dimension-$4$ in our case) usually requires an infinite number of the
counterterms to compute a physical quantity. This precludes one from the
direct use of the RG-equations for large logarithmic contributions to the
physical quantities. A generalization of the RG-equations for the summation of leading logs
in a wide class of non-renormalizable theories (deformed by irrelevant operators) was
developed
in Refs.~\cite{Kivel:2008mf,Kivel:2009az,Koschinski:2010mr,Polyakov:2010pt}.
An adaptation of this approach for the special case of 2D non-renormalizable QFTs was considered in Refs.~\cite{Polyakov:2018rdp,Linzen:2018pvj}.
It is worth mentioning that
similar equations, although in a different context,
were derived in Refs.~\cite{Kazakov:2016wrp,Borlakov:2016mwp,Borlakov:2017vra,Kazakov:2019wce,Kazakov:2020xbo,Kazakov:2020mfp}.

Generically, a physical observable (say $2\to 2$ scattering amplitude) in the theory (\ref{LagrangianBissextile}) at the  $(n+m-1)$-th loop order with
$n$ vertices $\sim g_1, g_2$  and $m$ vertices $\sim 1/F^2$ at low energies acquires a contribution of large LLs of the type:
\begin{align}
\sim \left(\frac{E^2}{4\pi G}\right)^n\left(\frac{1}{4\pi F^2}\right)^m\omega_{n,m}\ln^{n+m-1}\left(\frac{\mu^2}{E^2}\right).\label{M(E)}
\end{align}
Here $E$ is the typical energy scale for a physics observable under consideration;
$\mu$ is an arbitrary mass scale introduced
to perform ultraviolet (UV) renormalization.
The numerical coefficients $\omega_{n,m}$ depend only on the basic parameters of
the theory, like the dimension of the field manifold $N$.
It is remarkable that the constants $g_1$ and $g_2$, which describe the general dimension-$4$ irrelevant deformation of the 2D $\sigma$-model,
enter the answer only in the combination $2g_1+g_2 =1/G$.
Equivalently, it is stated that the coefficients $\omega_{n,m}$ are
independent of the dimensionless combination $\lambda G$. The constant $\lambda$ which describes the $T\bar T$ deformation of the theory
(\ref{LagrangianBissextile2})
does not enter the LLs contributions. In the following sections we  demonstrate
explicitly the above statements within the theory
(\ref{LagrangianBissextile2})
computing the leading log coefficients
$\omega_{n,m}$
for
$2 \to 2$
scattering amplitude and performing the all-order summation of the LL-contributions.

\section{Scattering amplitude in deformed theory}

The goal of the current paper is to study the $2 \to 2$ particle scattering of the $O(N+1)$-symmetric theory (\ref{LagrangianBissextile2}):
\begin{align}
\Phi_a(p_1)+\Phi_b(p_2)\rightarrow\Phi_c(p_3)+\Phi_d(p_4)
\end{align}	
in the leading logarithmic approximation.
Here the $\Phi$-fields are the local coordinates of the theory
(\ref{LagrangianBissextile2})
related to the $\vec{n}$-fields through the standard parametrization
\begin{align}
\vec{n}=\left(\Phi^1,\ldots\Phi^{N},\sqrt{ {F^2}-\vec{\Phi^2}}\right),
\end{align}
where $F$ is the radius of the field manifold
$S^N=O(N+1)/O(N)$.
The $2\to2$ scattering amplitude
is a function of the Mandelstam variables
\begin{align}
s=\left(p_{1}+p_{2}\right)^{2} ; \quad t=\left(p_{1}-p_{4}\right)^{2} ; \quad u=\left(p_{1}-p_{3}\right)^{2}.
\end{align}
The $O(N)$ group (to which we refer as the isospin) indices
are projected over the irreducible representations as
\begin{align}
\mathcal{M}_{a b c d}(s, t, u)=\sum_{I=0}^{2} P_{a b c d}^{I} \mathcal{M}^{I}(s, t, u).
\end{align}
Here $\mathcal{M}^{I}$ are the invariant amplitudes satisfying
the following crossing symmetry relations:
\be
&&
\mathcal{M}^{I}(s, t, u)=C_{s t}^{I J} \mathcal{M}^{J}(t, s, u); \nn \\ &&
\mathcal{M}^{I}(s, t, u)=C_{s u}^{I J} \mathcal{M}^{J}(u, t, s); \nn \\ &&
\mathcal{M}^{I}(s, t, u)=C_{t u}^{I J} \mathcal{M}^{J}(s, u, t).
\ee
The definitions and properties of the projection operators $P_{a b c d}^{I}$ and the crossing matrices $C_{s t}^{I J}$, $C_{s u}^{I J}$ and $C_{t u}^{I J}$ can be found in Appendix~\ref{Group}.

In 2D it is possible to further simplify  the scattering amplitude by introducing the reflection ($R$) and transmission ($T$) amplitudes
\be
\mathcal{M}^{I,T}(s)=\mathcal{M}^{I}(s,t=0,u=-s); \nn \\
\mathcal{M}^{I,R}(s)=\mathcal{M}^{I}(s,t=-s,u=0).
\ee
From a general dimensional analysis, the leading logarithmic contributions to the $2\to 2$ transmission and reflection scattering amplitudes can be parametrized as follows:
\be
\label{eq:LLamp}
\mathcal{M}(s)^{I, T / R}&=&4 \pi s \sum_{n, m=0,n+m\geq1}^{+\infty}\left(\frac{s}{4 \pi G}\right)^{n}\left(\frac{1}{4 \pi F^{2}}\right)^{m} \omega_{n, m}^{I, T / R} \ln^{n+m-1} \left(\frac{\mu^{2}}{s}\right)\\
&+&\mathcal{O}(N L L).
\nonumber
\ee
Here $\mathcal{O}(N L L)$ stands for the contributions of the next-to-leading logs.
From general dimensional considerations the LL coefficients are
functions of the group order parameter $N$ and of the dimensionless combination of couplings $\lambda G$ of the theory (\ref{LagrangianBissextile2}).
For example, the tree-level LL-coefficients (corresponding to $n+m=1$, zero number of loops) have the form:
\be
\omega_{1,0}^{0, T}&=\omega_{1,0}^{0, R}&=- 4\lambda G+ (N+3); \nn \\
 \omega_{1,0}^{1, T}&=-\omega_{1,0}^{0, R}&=-4\lambda G+1; \nn \\
 \omega_{1,0}^{2, T}&=\omega_{1,0}^{2, R}&=-4 \lambda G+3; \nn \\
\omega_{0,1}^{0, T}&=\omega_{0,1}^{0, R}&=N-1; \nn  \\ \omega_{0,1}^{1, T}&=-\omega_{0,1}^{1, R}&=1; \nn \\ \omega_{0,1}^{2, T}&=\omega_{0,1}^{2, R}&=-1.
\label{eq:inclast}
\ee
Indeed, we observe that  the tree-level coefficients  depend on both $N$ and $\lambda G$. However, the explicit calculation of the one-loop Feynman graphs
demonstrates that the corresponding coefficients $\omega_{n,m}$ (with $n+m=2$) are independent of $\lambda G$. For higher loops the $\lambda G$-independence of
$\omega_{n,m}$ (with $n+m\ge2$)
follows from the recursion equations we discuss below.

Following the derivation detailed in Section~V of Ref.~\cite{Koschinski:2010mr}
for an arbitrary even spacetime dimension $>2$,
and in Appendix A of \cite{Linzen:2018pvj} for 2D QFTs,
we obtain the system of recurrence relations for the
$\omega^{I,T/R}_{n,m}$
coefficients:
\be
\label{eq:RR}
 \omega_{n, m}^{I, T} &=\frac{1}{n+m-1} \sum_{i=0}^{n} \sum_{j=0}^{m} \frac{1}{2}\left(\delta^{I J}-(-1)^{n} C_{s u}^{I J}\right)\left[\omega_{i, j}^{J, T} \omega_{n-i, m-j}^{J, T}+\omega_{i, j}^{J, R}  \omega_{n-i, m-j}^{J, R}\right]; \nn \\ \omega_{n, m}^{I, R} &=\frac{1}{n+m-1} \sum_{i=0}^{n} \sum_{j=0}^{m} \frac{1}{2}\left(\delta^{I J}-(-1)^{n} C_{s t}^{I J}\right)\left[\omega_{i, j}^{J, T} \omega_{n-i, m-j}^{J, R}+\omega_{i, j}^{J, R} \omega_{n-i, m-j}^{J, T}\right].
\ee
It is important to note, that the repeated isospin indices of the $\omega$-coefficients in
(\ref{eq:RR})
do not imply  summation. The initial conditions for the recurrence
relations are given by the tree-level results summarized in Eq.~(\ref{eq:inclast}).
A remarkable property of above recurrence relations is that the initial
conditions
(\ref{eq:inclast}) with  terms $\sim \lambda G$ only provide zero solutions for $\omega_{n,m}$. Therefore, we confirm that the LL coefficients in  expansion (\ref{eq:LLamp}) are independent of the parameter $\lambda$ of $T\bar T$ perturbation. This finding reflects the integrability of
the $T\bar T$ deformation. We can speculate here that the search of  the initial conditions providing
``zero solutions'' of the recurrence relations of the
type (\ref{eq:RR}) in various deformed theories can be used to identify other than $T\bar T$ integrable deformations of the theory.

The recurrence relations (\ref{eq:RR}) can be further simplified
employing the Bose symmetry:
\be
\omega_{n,m}^{I,T}=(-1)^{I}\omega_{n,m}^{I,R};
\ee
and the crossing symmetry:
\be
\omega_{n, m}^{I,T}&=  -C_{s u}^{I J} \omega_{n, m}^{J,T}\quad &\text { for even } n; \nn  \\
\omega_{n, m}^{I,T}&=C_{s u}^{I J} \omega_{n, m}^{J,T}\quad &\text { for odd } n.
\ee
This eventually leads  to the relations
\begin{align}
\frac{\omega^{1,T}_{n,m}}{\omega^{0,T}_{n,m}}&=
\begin{dcases}
\frac{1}{N-1},~n-{\rm even;}\\
-\frac{1}{N-1},~n-{\rm odd;}
\end{dcases} \nn \\
\frac{\omega^{2,T}_{n,m}}{\omega^{0,T}_{n,m}}&=
\begin{dcases}
-\frac{1}{N-1},~n-{\rm even;}\\
\frac{N-2}{(N-1)(N+2)},~n-{\rm odd.}
\end{dcases}
\end{align}
Using the above relations we manage to close the recurrence relation
for the transmission amplitude involving only the $I=0$ isospin channel:
\be
\omega_{n, m}^{0, T}=\frac{1}{m+n-1} \sum_{k=0}^{n} \sum_{l=0}^{m}\left(A_{0}+(-1)^{n} A_{1}+(-1)^{k} A_{2}\right) \omega_{k, l}^{0, T} \omega_{n-k, m-l}^{0, T}
\label{RROmegaT}
\ee
with the coefficients
\be
&&
A_{0}=1+\frac{1}{(N+2)(N-1)}, \quad A_{1}=-\frac{N+1}{(N+2)(N-1)}, \quad A_{2}=-\frac{2}{(N+2)(N-1)}.
\nn \\ &&
\label{eq:Acoeffs}
\ee
We introduce the generating function
for the LL coefficients $\omega_{n,m}^{0,T}$:
\be
\label{eq:omdef}
\Omega(z,w)=\sum_{n,m=0,n+m\geq2}\omega_{n,m}^{0,T}z^{n+m-1}w^m.
\ee
Note that the function $\Omega(z,w)$ contains only the loop contributions because of the constraint of the summation $n+m\geq2$;
therefore, we will call
the function $\Omega(z,w)$ as the ``loop function''.
We can express all LL amplitudes (different isospins, transmission, reflection) in terms of the universal function $\Omega(z,w)$.
For example, the isospin-$0$ transmission amplitude in the LL approximation has the form%
\footnote{We report a missing factor $\frac{1}{2}$ in the arguments of the $\Omega$-functions in the Eq. (2.21) of Ref.~\cite{Linzen:2018pvj}.}:
\be
\nonumber
\mathcal{M}^{0,T}(s)=\frac{s}{F^2}(N-1)+s^2\left(-4\lambda+\frac{1}{G} (N+3)\right)+\frac{s^2}{G}\Omega\left(\frac{s}{4\pi G}\ln\left(\frac{\mu^2}{s}\right),\frac{G^2}{sF^2}\right).\\
\label{eq:amplviaom}
\ee
All other isospin amplitudes (transmission and reflection) can  also be expressed
in terms of the function $\Omega(z,w)$, corresponding expressions are
summarized in Appendix~\ref{OmegavsA}.

We can further simplify the recurrence relation (\ref{RROmegaT}) by introducing the
rescaled coefficients $f_{n,m}$:
\be
\label{eq:rescalingomtof}
\omega_{n, m}^{0, T}=f_{n, m}\left(\frac{(N-1)(N+2)}{N} \right)^{n}(N-1)^{m}
\ee
leading to the master recurrence relation:
\be
f_{n,m}=\frac{1}{n+m-1} \sum_{k=0}^{n} \sum_{l=0}^{m}\left(A_{0}+(-1)^{n} A_{1}+(-1)^{k} A_{2}\right)f_{k, l} f_{n-k, m-l},\label{MasterRR}
\ee
with the initial conditions:
\be
\label{eq:inco}
f_{0,0}=0, \ f_{1,0}=1,\ f_{0,1}=1.
\ee
Below, we reduce this discrete equation to the non-linear differential equations
for the generating function (loop function) (\ref{eq:omdef})
and present some of its exact solutions.

\section{Generalizing RG-equations: non-linear differential equations for the LL-amplitude}
\label{DiffEq}

The recurrence relation (\ref{MasterRR})
can be reduced to the non-linear differential equation which
takes the form of the equation of motion of the ``equivalent mechanical system''.
It corresponds to a ``particle'' moving in 1D
in a potential with the form depending on the dimension of the field manifold $N$.
The details of the reduction to the ``equivalent mechanical systems'' in the limit
$G\to \infty$ ($m=0$ case in Eq.~ (\ref{MasterRR})) are
presented in Ref.~\cite{Linzen:2018pvj}.
Here we  review only the main steps of the derivation and provide the result
for a more complicated case of the recurrence relations
(\ref{MasterRR}).

In order to reduce the recurrence relations (\ref{MasterRR}) to the differential equation 
we introduce
the generating function:
\be
\Phi(z,w)=\sum_{n,m=0}^{\infty} f_{n, m} z^{n+m-1} w^m.
\ee
From the initial conditions (\ref{eq:inco}) we obtain:
\be
\label{eq:incophi}
\Phi(0,w)=1+w.
\ee
We can express the generating function for the loop amplitudes $\Omega(z,w)$ through the
function $\Phi(z,w)$. Indeed, the relation
(\ref{eq:rescalingomtof}) and the definition of the loop function (\ref{eq:omdef}) imply:
\be
\label{eq:omthroughphi}
\Omega(z,w)&= &\frac{(N+2)(N-1)}{N} \left(\Phi(z',w')-1-w'\right), \ {\rm with}\\
\nonumber
z'&=&\frac{(N+2)(N-1)}{N}\ z; \ \ \ w'=\frac{N}{N+2} w.
\ee

The next step is to split the generating function $\Phi(z,w)$ into the sum of functions
with specific symmetry with respect to $z\to -z$ and $w \to -w$:
\be
\label{eq:split}
\Phi(z,w)=\Phi_{++}(z,w)+\Phi_{+-}(z,w)+\Phi_{-+}(z,w)+\Phi_{--}(z,w).
\ee
The subscripts in
(\ref{eq:split})
refer to evenness/oddness of the corresponding function with respect to the arguments
$z$ and  $w$.
{\it E.g.}
$\Phi_{+-}(z,w)$
is even in the argument
$z$ and is odd in the argument
$w$. From (\ref{eq:incophi}) and the symmetry properties
of the functions $\Phi_{\pm \pm}(z,w)$ 
one can conclude that:
\be
\Phi_{-+}(0,w)=\Phi_{--}(0,w)=0, \ {\rm and}\ \Phi_{++}(0,w)=1, \ \Phi_{+-}(0,w)=w.
\label{Init_Cond_Phi}
\ee
Further, differentiating Eq.~(\ref{eq:split}) in $z$
and employing the recurrence relation (\ref{MasterRR})
one can obtain the system of coupled differential equations for the functions
$\Phi_{\pm \pm}(z,w)$ with the initial conditions provided by
(\ref{Init_Cond_Phi}).
Collecting terms with the same symmetry properties under $z\to -z$ and $w\to -w$
we obtain the system of coupled non-linear differential equations for the functions
\be
u(z,w)=\Phi_{++}(z,w)+\Phi_{--}(z,w), \ {\rm and}\  v(z,w)=\Phi_{+-}(z,w)+\Phi_{-+}(z,w).
\ee
The resulting differential equations have exactly the same form as those considered in
Ref.~\cite{Linzen:2018pvj} (see Eq.~(3.6) of that paper):
\be
\begin{cases}
\frac{\partial}{\partial z}v(z,w)= (A_0+A_1-A_2) u^2(z,w)+ (A_0+A_1+A_2) v^2(z,w);
\\
\frac{\partial}{\partial z}u(z,w)=2 (A_0-A_1) u(z,w) v(z,w);
\end{cases}
\label{DifUr}
\ee
but with the different initial conditions:
\be
u(0,w)=1; \ \ \ v(0,w)=w.
\ee
Obviously:
\be
\Phi(z,w)=u(z,w)+v(z,w).
\ee
In the limit of zero curvature of the field manifold ($F \to \infty$, or, equivalently, $w\to 0$) we obtain exactly the same equations as in Ref.~\cite{Linzen:2018pvj}.
For the more general system (\ref{DifUr}) we can repeat, with small modifications, the derivation of the equivalent mechanical system.
Here we provide only
the final result, the main calculations
steps
are well described in  Ref.~\cite{Linzen:2018pvj}
(see also Ref.~\cite{SokolovaDiplom}).
The generating loop function $\Omega(z,w)$ can be obtained in the form:
\be
\label{eq:omviaq}
\Omega(z,w)=\frac{(N+2)(N-1)}{N}\left(\frac{1}{q(z,w)}-1-\frac{N}{N+2}w\right)-\frac{N-1}{2N}\frac{\partial}{\partial z}\ln(q(z,w)),
\ee
where the function $q(z,w)$
can be formally viewed as the trajectory obtained from the equation of motion for a particle of the mass
\be
m=\frac{1}{2 N^2}
\label{Def_mass_eq_mech}
\ee
and the total energy $E=1$ in one dimension (along the coordinate $q$):
\be
\label{eq:mech_system}
\frac{m}{2} \dot q(t,w)^2  +(1-w^2)q(t,w)^\gamma  =1,
\ee
where
\be
\gamma=\frac{N+2}{N}
\label{Def_exponent_eq_mech}
\ee
is the exponent of the potential. The initial conditions  corresponding to
$t=0$
are the following:
\be
\label{eq:incomech}
q(0,w)=1; \ \ \ \dot{q}(0,w)=-2 N w.
\ee
As it must be in the limit $w\to 0$ (zero curvature limit of the field manifold ) the equivalent
mechanical system (\ref{eq:mech_system}) is reduced to that considered in
Ref.~\cite{Linzen:2018pvj}. In that paper a number of interesting solutions
were found. In the next Section we analyze in details the solutions
of the mechanical system
(\ref{eq:mech_system}), (\ref{eq:incomech})
with $w \neq 0$ and present the resulting LL scattering amplitudes.

\section{Approximate solutions for the LL-scattering-amplitude}
\label{Sol}

\subsection{Solution for pure $T\bar T$ deformed 2D $\sigma$-model}
\label{sec:RT}

The pure $T\bar T$ deformation of the 2D $O(N+1)$ $\sigma$-model corresponds
to the limit $G\to\infty$, see Eq.~(\ref{LagrangianBissextile2}).
In terms of the variables $z,w$ it corresponds to the limits:
$w\to \infty$, $z\to 0$, $w z\to y$
with fixed $y$. Performing this limit for the loop function
$\Omega(z,w)$
we obtain the loop function within the $T\bar T$ theory:
\be
\label{eq:omttbar}
\Omega^{T\bar T} (y) =\lim_{w\to \infty\atop z\to 0,  w z=y} \frac{1}w \Omega(z,w).
\ee
In terms of this function the LL amplitude%
\footnote{We consider only the transmission amplitude with the isospin
$I=0$;
results for other amplitudes can be obtained using the relations summarized in Appendix~\ref{OmegavsA}.}
takes the form:
\be
\label{eq:amplttbar}
\mathcal{M}(s) =\frac{s\ (N-1)}{F^2}- 4s^2\lambda+\frac{s\ (N-1)}{F^2} \Omega^{T\bar T} \left(\frac{1}{4 \pi F^2} \ln\left(\frac{\mu^2}{s}\right) \right).
\ee
The function $\Omega^{T\bar T} (y)$  can be obtained by solving the equivalent mechanical system (\ref{eq:mech_system}) in the limit
$w\to\infty$.  For this issue, we rescale the time variable $t=y/w$ and take the limit $w\to\infty$ in Eq.~(\ref{eq:mech_system}). This results
in the simplified equation:
\be
\frac{m}{2} \left( \frac{d}{dy} q(y)\right)^2 =q(y)^\gamma,
\ee
where $m$ and $\gamma$ are defined  in (\ref{Def_mass_eq_mech}), (\ref{Def_exponent_eq_mech}).
Taking into account the initial conditions
(\ref{eq:incomech})
this equation can be solved with the result:
\be
\label{eq:zeroorder}
q(y)=\left(1-\sqrt{\frac 2m} \left(1-\frac\gamma 2\right) y\right)^{\frac{2}{2-\gamma}}.
\ee
Having the general solution we can with help of Eq.~(\ref{eq:omviaq}) obtain  the loop function in the
$T\bar T$ theory as:
\be
\label{eq:omttbar_1}
\Omega^{T\bar T} (y)=(N-1)\frac{(N-2) y}{1-(N-2) y},
\ee
which with help of Eq.~(\ref{eq:amplttbar}) leads to the following LL resummed
scattering amplitude for arbitrary
$N$:
\be
\label{eq:betaf}
\mathcal{M}(s)&=&-4s^2\lambda+\frac{s}{F^2}(N-1) \left(\frac{1}{1-\frac{(N-2)}{4 \pi F^2} \ln\left(\frac{\mu^2}{s}\right)}\right)\\
\nonumber
&=&-4s^2\ \lambda+\frac{s}{F^2(s)}(N-1).
\ee
We verify that the LL resummed amplitude in the $T\bar T$ deformed theory
coincides with the tree-level amplitude, in which
the coupling constant $F$ is replaced by the corresponding running coupling constant $F(s)$.
This demonstrates that our generalization of the
RG equations (RGE) is reduced to the usual RGE for the case of
renormalizable theories.
However, there is a subtlety here.  The $\beta$-function\footnote{The $\beta$-function is defined as $d \left(\frac{1}{F^2}\right)/d(\ln(\mu^2)) $.}
corresponding to the running of the coupling in Eq.~(\ref{eq:betaf}) is:
\be
\label{eq:betanawa}
\beta\left(\frac{1}{F^2}\right)=\frac{N-2}{4 \pi} \frac{1}{F^4}+O\left(\frac{1}{F^6}\right),
\ee
whereas the $\beta$-function obtained from UV renormalization of the $O(N+1)$ 2D $\sigma$-model \cite{Polyakov:1975rr}
contains the factor $(N-1)$ instead of $(N-2)$ in above equation. The reason for this
discrepancy is that the 2D $\sigma$-model
is suffering from the infrared (IR) divergencies that can be regularized introducing
the mass term to the action
(\ref{LagrangianBissextile2}).
In the theory with the mass term the scattering amplitude
obtains logarithmic corrections of the type $\sim \ln(\mu^2/s)$ and $\sim \ln(\mu^2/m_{g})$, where $m_g$ is the
mass of (pseudo)Goldstone boson. The method of RG equations sums up {\it both} types of the corrections, whereas our method based on analyticity
of the amplitude in the variable $s$ performs the summations only of the first type of LLs.
It was shown by the explicit one-loop calculation in Ref.~\cite{LinzenDiplom}
that the contribution of the mass logs to the scattering
amplitude corresponds to the shift of the
coefficient of the $\beta$-function in Eq.~(\ref{eq:betanawa}) by one unity, thus
bringing it to the correct value. This study will be published elsewhere.

\subsection{Expansion around the $T\bar T$ deformation of the 2D $\sigma$ model}
\label{sec:expansion}

In the previous Section we considered the exact limit of
$G\to\infty$ corresponding to the pure $T\bar T$ theory. Now we can systematically
compute $1/G$ corrections to the amplitude. This expansion corresponds to the kinematical domain
\be
\frac{1}{4\pi F^2} \ln\left(\frac{\mu^2}{s}\right)\gg
\frac{s}{4\pi G} \ln\left(\frac{\mu^2}{s}\right),
\ee
in which we sum up exactly the logs proportional
$\sim 1/F^2$
and perform perturbative expansion to the finite $n$-th order in logs $\sim 1/G$.
Therefore, the corresponding expansion around the $T\bar T$ theory can be obtained
by expanding the loop function in small $z$ and large $w$ (with fixed $w z=y$):
\be
\label{eq:expamsioninz}
\frac{z}{y}\Omega\left(z,\frac{y}{z}\right) = \Omega^{T\bar T}(y)+ \Omega_1^{T\bar T}(y) z +\Omega_2^{T\bar T}(y) z^2+\ldots,
\ee
where the zeroth order amplitude $ \Omega^{T\bar T}(y)$
(\ref{eq:omttbar_1}) was computed in the previous Section.
The higher order functions $\Omega_n^{T\bar T}(y)$ describe the corrections due to the non-integrable deformation of the theory.
The functions $ \Omega_n^{T\bar T}(y)$ can be found by solving the mechanical system perturbatively  at large $w$. The mechanical
equations after the rescaling of time $t=y/w$ have the form:
\be
\frac{m}{2} \left( \frac{d}{dy} q(y)\right)^2 =q(y)^\gamma+\frac{1}{w^2} (1-q(y)^\gamma),\  {\rm with}\ \frac{d}{dy} q(0)=-2 N.
\ee
The above equation can be solved by perturbation in small $1/w$ starting from the zeroth order solution (\ref{eq:zeroorder}).
Alternatively, one can obtain the  recurrent differential equations for $ \Omega_n^{T\bar T}(y)$ directly from (\ref{MasterRR}). The corresponding differential
equations are linear and can be easily integrated. Below we provide the final result for the corrections.

The first correction has the following loop function:
\be
\frac{\Omega_1^{T\bar T}(y)}{N-1} =\frac{(N+2)}{N}\frac1y\left(\frac{1}{(1-(N-2) y)^{\frac{2N}{N-2}}}-1\right).
\ee
The higher loop functions (for $n\ge 2$) are expressed as:
\be
\frac{\Omega_n^{T\bar T}(y)}{N-1}=\frac{1}{y^n} \Xi_n((N-1)y) \left[\frac{(N+2)}{N}\right]^n,
\ee
where $\Xi_n(y)$ can be computed iteratively as the integral ($n\ge 2$):
\be
\nonumber
\Xi_n(y)= \frac{1}{\left(1-\beta_0 y\right)^{\frac{2 \beta_n}{\beta_0}}} \int_0^y d\bar y\ \left[1-\beta_0 \bar y\right]^{\frac{2 \beta_n}{\beta_0}}
\sum_{k=1}^{n-1} A(n,k) \Xi_{n-k}(\bar y)\ \Xi_{k}(\bar y),\\
\label{eq:Xirec}
\ee
with the starting function:
\be
\Xi_1(y)&=&\frac{1}{\left(1-\beta_0 y\right)^{\frac{2 \beta_1}{\beta_0}}}.
\ee
The coefficient function $A(n,k)$ is defined as:
\be
A(n,k)=\left(A_{0}+(-1)^{n} A_{1}+(-1)^{k} A_{2}\right),
\ee
with coefficients $A_{0,1,2}$ given in Eq.~(\ref{eq:Acoeffs}) in terms of the group order parameter $N$.
Eventually, the coefficients $\beta_n$ in Eq.~(\ref{eq:Xirec}) are defined as:
\be
\beta_n=\frac 12 \left(A(n,0)+A(n,n)\right).
\ee
This gives:
\be
\beta_{\rm even}=\frac{N-2}{N-1}; \ \ \  \beta_{\rm odd}=\frac{N}{N-1}.
\ee
Note that the complete loop function for $N=-2$ does not have corrections in $1/G$, {\it i.e.} the corresponding
amplitude coincides with the amplitude in $T\bar T$ theory. It means that the most general deformation of the
2D $O(N+1)$-symmetric $\sigma$-model (\ref{LagrangianBissextile2}) remains integrable for $N=-2$.
In what follows we will also
obtain this result 
from the exact summation of the LLs in the limit $N\to -2$.

The explicit form of the amplitude%
\footnote{We consider only the transmission amplitude with isospin $I=0$, results for other amplitudes can be
obtained using relations in Appendix~\ref{OmegavsA}.}
to the first order in the perturbation around the $T\bar T$ theory is the following:
\be
\label{eq:betaf1stcorr}
\mathcal{M}(s)&=&-4s^2\ \lambda+\frac{s}{F^2}(N-1) \left(\frac{1}{1-\frac{(N-2)}{4 \pi F^2} \ln\left(\frac{\mu^2}{s}\right)}\right) \\&+&
\nonumber
\frac{s^2}{G}\left[(N+3)
 +  \frac{(N-1)(N+2)}{N} \left(\frac{1}{\left[1-\frac{(N-2)}{4 \pi F^2} \ln\left(\frac{\mu^2}{s}\right)\right]^{\frac{2N}{N+2}}}-1\right) \right]+O\left(\frac{1}{G^2}\right)
\ee
Here the first line corresponds to the amplitude in the pure $T \bar T$ theory
(\ref{eq:betaf}).

\subsection{$1/N$ expansion}

The $O(N+1)$-symmetric model in the large-$N$ limit in an arbitrary space-time
dimension is equivalent to a renormalizable field theory, see {\it e.g.}
discussion in \cite{ANV}. Our  generalization of the RG equations
(\ref{DifUr}) in the
$N\to\infty$
limit are reduced to the following simple differential equation:
\be
\label{eq:RGlike}
\frac{\partial}{\partial z} \left[\frac1N\Omega(z,w)+ (1+w)\right]=N\ \left[\frac1N\Omega(z,w)+ (1+w)\right]^2,{\rm with}\ \Omega(0,w)=0,
\ee
which has the form of the usual one-loop RG equation with the one-loop $\beta$-function coefficient equal to $N$.
The corresponding equation can be easily solved with the result:
 \be
\Omega(z,w)=\frac{N(1+w)}{1-N(1+w) z} -N(1+w).
\ee
This solution has a typical analytical structure for a solution of one-loop RG equations
-- it contains a single Landau pole.

\section{Exact solutions for the LL-scattering-amplitude}
\label{Sol2}

In the previous Section we computed the LL amplitudes in the theory
(\ref{LagrangianBissextile2})
employing several types of approximations. Now we discuss the
results of the exact summation of the leading logs.

\subsection{Amplitude at the special kinematic point}
\label{sec:specialpoint}

The coupling constant ratio $G/F^2$ has the mass dimension 2 and
provides the natural mass scale for the theory. In this subsection we consider
the LL-resummed amplitude at the special kinematic point:
\be
\label{eq:specialpoint}
s=s_0=\frac{|G|}{F^2}\to 0,
\ee
which corresponds to $w=\pm 1$, where the sign coincides with the sign of $G$.
The limit (\ref{eq:specialpoint})
ensures that we stay in the low energy domain
where the log contributions which we sum up are large.
Physically, the limit (\ref{eq:specialpoint}) corresponds to either
the case of the strong deviation from the $T\bar T$ deformation, or,
equivalently, the case of the field manifold with a small curvature ($F\gg 1$).

For  $w=\pm 1$ the mechanical system (\ref{eq:mech_system}) describes the motion with a
constant velocity:
\be
q(t) = 1\mp2 N t.
\ee
With help of Eq.~(\ref{eq:omviaq}) we obtain the loop function
with a single pole in the $z$-plane:
\be
\Omega(z,  +1)&=&\frac{4 (N^2-1) z}{1-2 N z}; \nn \\
\Omega(z,  -1)&=&\frac{4 (N-1)z}{1+2 N z}.
\label{Single_pole_sol}
\ee
Employing Eq.~(\ref{eq:amplviaom}) we establish
the following expression
for the  LL-approximation amplitude%
\footnote{$\mathcal{M}^{\rm loops}$ stands for the contribution into
the LL amplitude
starting from the one-loop order.
We consider only the transmission amplitude with isospin $I=0$. Other amplitudes can be
obtained using the relations summarized in Appendix~\ref{OmegavsA}.}:
\be
\mathcal{M}^{\rm loops}(s_0)&=&\frac{G}{F^4}\frac{4 (N^2-1)}{1-\frac{2 N}{4\pi F^2}\ln\left(\frac{\mu^2 F^2}{|G|}\right)}\cdot \frac{1}{4\pi F^2}\ln\left(\frac{\mu^2 F^2}{|G|}\right), \ {\rm for}\ G>0; \nn \\
\mathcal{M}^{\rm loops}(s_0)&=&\frac{|G|}{F^4}\frac{4 (N-1)}{1-\frac{2 N}{4\pi F^2}\ln\left(\frac{\mu^2 F^2}{|G|}\right)}\cdot
\frac{1}{4\pi F^2}\ln\left(\frac{\mu^2 F^2}{|G|}\right), \ {\rm for}\ G<0.
\ee
Note the remarkable property of the above amplitudes: they possess a pole if the coupling constants of the theory are related by:
\be
\frac{G}{F^2}=\mu^2 e^{-4 \pi F^2/(2 N)}.
\ee

Since the amplitude at the point $s=s_0$ is exactly solvable one can easily construct the systematic expansion of the amplitude
around this point. As an example we present here the first correction to the amplitude for $s=s_0+\delta s$ with $\delta s/s_0\ll 1$
(we give result for $G>0$):
\be
\mathcal{M}^{\rm loops}(s_0+\delta s)&=&\frac{G}{F^4}\frac{4 (N^2-1)}{1-\frac{2 N}{4\pi F^2}\ln\left(\frac{\mu^2 F^2}{|G|}\right)}\cdot \frac{1}{4\pi F^2}\ln\left(\frac{\mu^2 F^2}{|G|}\right)\\
\nonumber
&-&\delta s\ \frac{N (N-1)}{ F^2}
\frac{\left[1-\frac{ N}{4\pi F^2}\ln\left(\frac{\mu^2 F^2}{|G|}\right)\right]}{\left[1-\frac{2 N}{4\pi F^2}\ln\left(\frac{\mu^2 F^2}{|G|}\right)\right]^2 }
 \cdot \frac{1}{4\pi F^2}\ln\left(\frac{\mu^2 F^2}{|G|}\right).
\ee

\subsection{Solutions for $N=2$: infinite number of equidistant poles }

The theory (\ref{LagrangianBissextile2}) for $N=2$ is of special interest as it can have many applications for descriptions of
physical systems like magnets, also this theory (in Euclidean space) possesses the conserved topological charge.

In Ref.~\cite{Linzen:2018pvj} the loop function in the zero curvature limit
($F\to \infty$, or $w=0$)
was found to have a remarkably
simple, but non-trivial form:
\be
\Omega(z,w=0)=-2 +\frac{2}{\cos(4 z)} +{\rm tg}(4 z),
\ee
exhibiting infinite number of equidistantly 
distributed poles in the variable $z$.  

For the general -- non-zero curvature -- case the loop function can be also easily found because the equivalent mechanical system (\ref{eq:mech_system})
corresponds to the harmonic oscillator. Note, however, that for the general case the form of the harmonic potential depends on
$w$,
and the motion of the equivalent mechanical system is of a qualitatively different nature for
$|w|<1$
and
$|w|>1$.
As a 
consequence
the corresponding solution for the loop function has essentially different pole
structure in the $z$-plane for
$|w|<1$
and
$|w|>1$.
\bi
\item
For $|w|<1$ we define
$w=\sin(\alpha)$ with $\alpha\in [-\pi/2,\pi/2]$. The loop function has the following form:
\be
\Omega(z,w)=\frac{2\cos(\alpha)}{\cos(\alpha+4 z\cos(\alpha))} +\frac{\sin(4 z\cos(\alpha))}{\cos(\alpha+4 z\cos(\alpha))}-2.
\ee
We see that in this case the loop function has equidistantly
distributed poles along  the real axis of the variable $z$.
\item
For $|w|>1$ we define $w=\pm{\rm ch}(\alpha)$ with $\alpha\in [0,\infty]$. The corresponding loop function is:
\be
\Omega(z,w)=\frac{2{\rm sh}(\alpha)}{{\rm sh}(\alpha\mp 4 z\ {\rm sh}(\alpha))} +\frac{{\rm sh}(4 z\ {\rm sh}(\alpha))}{{\rm sh}(\alpha\mp 4 z\ {\rm sh}(\alpha))}-2.
\ee
In this case the equidistantly
distributed poles
lie on the imaginary axis in the complex $z$ plane.
\ei
In both cases the distance between poles is increasing as $w$ approaches $1$. Therefore, all poles move to the infinity in this limit,
only one pole at $z=1/4$ remains at a finite position
for $w=1$. This corresponds to the $N=2$ case for the one-pole solution
(\ref{Single_pole_sol}) presented in Section~\ref{sec:specialpoint}.

\subsection{Other solutions in terms of elementary functions}

Below we list the exact solutions for the loop function $\Omega(z,w)$ in terms of elementary functions.

\subsubsection{$N=-2$}

In Ref.~\cite{Linzen:2018pvj} it was shown that the LLs are absent in the theory with $N=-2$ 
for $F=\infty$ (deformation of the free field theory).
In the case of the field manifold with non-zero curvature ($1/F\neq 0$)
the
equivalent mechanical system (\ref{eq:mech_system}) for $N=-2$ corresponds to a motion with constant velocity.
The general solution reads:
\be
q(t)=1+ 4 w t.
\ee
The corresponding loop function has the form:
\be
\Omega(z,w)=\frac{12 w^2 z}{1+4 w z}.
\ee
Such loop function corresponds to the LL amplitude of the form:
\be
\label{eq:amplN-2}
\mathcal{M}^{\rm loops}(s)&=& \frac{3 s}{ F^2 }\left(1- \frac{1}{1+\frac{4}{4\pi F^2} \ln\left(\frac{\mu^2 }{s}\right)}\right).
\ee
We conclude that the LL amplitude
for $N=-2$
is indeed independent of $G$ confirming
the findings of Section~\ref{sec:expansion}, where
the perturbation around the $T\bar T$ theory was considered.
The exact summation of the LLs for the amplitude gives the amplitude
of the $T\bar T$ theory in which the LLs can be absorbed
into the running coupling constant. Indeed, the above expression for the LL amplitude can be put into the tree-level form replacing the coupling constant $F$
by the corresponding running coupling constant, as discussed in Section~\ref{sec:RT}. The amplitude (\ref{eq:amplN-2}) possesses an UV Landau pole; and
hence the theory (\ref{LagrangianBissextile2})  is IR asymptotically free for $N=-2$. We can speculate that this feature of the theory at $N=-2$ is related
to the fact that the $O(N+1)$-symmetric theory at negative $N$ is equivalent to a fermionic theory with $N_{\rm ferm}=-N/2$ Grassmann fields.
In the corresponding fermionic theory with $N_{\rm ferm}=1$ all operators of dimension 4 are identically zero due to the Grassmann nature of fields.
This correspondence will be elaborated elsewhere.

\subsubsection{$N\to 0$}

Exact summation of all-loop logs can also be performed for the theory
(\ref{LagrangianBissextile2}) in the limit $N\to 0$.
In this limit the equivalent mechanical system
(\ref{eq:mech_system}) can be easily solved
providing the result:
\be
q(t,w)=\frac{1}{\left[ w\ {\rm sh}(2 t)+{\rm ch}(2 t)\right]^N},
\ee
where the leading in $N\to 0$ result is shown. Further with help of Eq.~(\ref{eq:omviaq}) we obtain the corresponding loop function:
\be
\Omega(z,w)=-2 \ln \left[{\rm ch}(2 z)+w\ {\rm sh}(2 z)\right]-\frac{(1-w^2)\ {\rm th}(2z)}{1+w\ {\rm th}(2z)}.
\ee
We note that, in contrast to previously considered exact solution, the position of the singularities in the variable $z$
does not change qualitatively when the variable $w$ crosses unity.
This observation is in accordance with the analysis of Section~\ref{sec:specialpoint}: the exact solution for the special point
$w=1$ obtained in this section corresponds to non-singular loop function  for $N=0$.

\subsection{A solution in terms of elliptic functions}

As was shown in Ref.~\cite{Linzen:2018pvj},
in the limit of zero curvature of the field manifold
($1/F^2\to 0$, or, equivalently, $w=0$)
the equivalent mechanical system
(\ref{eq:mech_system})
possesses solutions expressed in terms of elliptic (meromorphic, doubly-periodic)
functions, leading to scattering amplitudes with a doubly-periodic structure of poles.
Here we argue that this interesting class of
solutions extends to the case of a more general equivalent mechanical system with $w \neq 0$.
Prominently the case with $N=1$ corresponds to $m=\frac{1}{2}$ and $\gamma=3$, leading to the differential equation
\be
\dot{q}(t,w)^2+4(1-w^2)q(t,w)^3-4=0;~q(0,w)=1;~\dot{q}(0,w)=-2w.
\label{N=1sys1}
\ee
\bi
\item
We first consider the case $0<w<1$.
To 
work out the solution of
(\ref{N=1sys1})
we define
$w=\sin(\alpha)$
and introduce the time-shifted equivalent mechanical system
$Q(t,w)=q(t-t_0,w)$
described by the following differential equation
\be
\frac{1}{4}\dot{Q}(t,w)^2+\cos^2(\alpha)Q(t,w)^3=1, ~Q(0,w)=\cos^{-\frac{2}{3}}(\alpha),~\dot{Q}(0,w)=0
\label{N=1sys2}
\ee
with zero initial velocity.
The solution of 
(\ref{N=1sys2})
can be obtained in terms of the well known Weierstra\ss\ elliptic
$\wp$-function with the invariants ${\rm g}_2=0,~{\rm g}_3=-4$:
\be
Q(t,w)=\frac{1}{\cos^{\frac{2}{3}}(\alpha)}
\frac{\wp\left(\cos^{\frac{2}{3}}(\alpha)t;0,-4\right)-2}{\wp\left(\cos^{\frac{2}{3}}(\alpha)t;0,-4\right)+1}.
\ee
This function has one real period
\be
\omega_3(w)=  \pi_3 \frac{1}{2^{\frac{1}{3}}} \cos^{-\frac{2}{3}}{\alpha}
\ee
and two complex periods
\be
\omega_1(w)=   \frac{1}{2} \frac{\omega_3(w)}{2} -i    \frac{\sqrt{3}}{2 } \frac{\omega_3(w)}{2}; \ \ \
\omega_2(w)=  \frac{1}{2} \frac{\omega_3(w)}{2} -i    \frac{\sqrt{3}}{2 } \frac{\omega_3(w)}{2},
\ee
where the constant
$\pi_3$
can be computed as
\be
\pi_3= 3 \int_{ 0}^{ 1} \frac{d y}{(1-y^3)^{\frac{2}{3}} }=
B \left (\frac{1}{3}, \, \frac{1}{3} \right)= \frac{\sqrt{3}}{2 \pi} \Gamma^3 \left( \frac{1}{3} \right),
\label{pi3}
\ee
where $B$ is the Euler beta function.

To obtain the solution
$q(t,w)=Q(t+t_0,w)$
for the initial mechanical system (\ref{N=1sys1})
one can employ the familiar addition theorem for the Weierstra\ss\  $\wp$-function
(see {\it e.g.} \cite{Pastras:2017wot})
\be
\wp(t+t_0)=-\wp(t)-\wp(t_0)+\frac{1}{4}\left(\frac{\wp'(t)-\wp'(t_0)}{\wp(t)-\wp(t_0)}\right)^2
\label{wpAddition}
\ee
together with the relations
\be
\wp(\cos^{\frac{2}{3}}(\alpha)t_0;0,-4)&=\frac{2+\cos^{\frac{2}{3}}(\alpha)}{1-\cos^{\frac{2}{3}}(\alpha)}; \nn \\
\wp'(\cos^{\frac{2}{3}}(\alpha)t_0;0,-4)&=\frac{-6\sin(\alpha)}{(\cos^{\frac{2}{3}}(\alpha)-1)^2}
\ee
established from the initial conditions of (\ref{N=1sys1}).
Note that for $w=0$ we indeed reproduce the familiar Flajolet's
\cite{Baher,Flajolet}
solution written in a form quoted in Refs. \cite{Linzen:2018pvj,Polyakov:2018rdp}.

\item For the case $w>1$ the substitution
$w={\rm ch}(\alpha)$ leads to the time-shifted solution
\be
Q(t,w)=\frac{-1}{{\rm sh}^{\frac{2}{3}}(\alpha)}\frac{\wp({\rm sh}^{\frac{2}{3}}(\alpha)t;0,-4)-2}{\wp({\rm sh}^{\frac{2}{3}}(\alpha)t;0,-4)+1}.
\ee
The corresponding real period is given by
\be
\omega_3(w)=  \pi_3 \frac{1}{2^{\frac{1}{3}}} {\rm sh}^{-\frac{2}{3}}{\alpha}.
\ee
The solution for the mechanical system (\ref{N=1sys1}) can again be obtained via (\ref{wpAddition}) and
\be
\wp({\rm sh}^{\frac{2}{3}}(\alpha)t_0;0,-4)&=\frac{2-{\rm sh}^{\frac{2}{3}}(\alpha)}{1+{\rm sh}^{\frac{2}{3}}(\alpha)}; \nn \\
\wp'({\rm sh}^{\frac{2}{3}}(\alpha)t_0;0,-4)&=\frac{6{\rm ch}(\alpha)}{({\rm sh}^{\frac{2}{3}}(\alpha)+1)^2}.
\ee
\ei

\begin{figure}[h]
\centering
\begin{subfigure}[b]{0.45\textwidth}
\includegraphics[scale=.45]{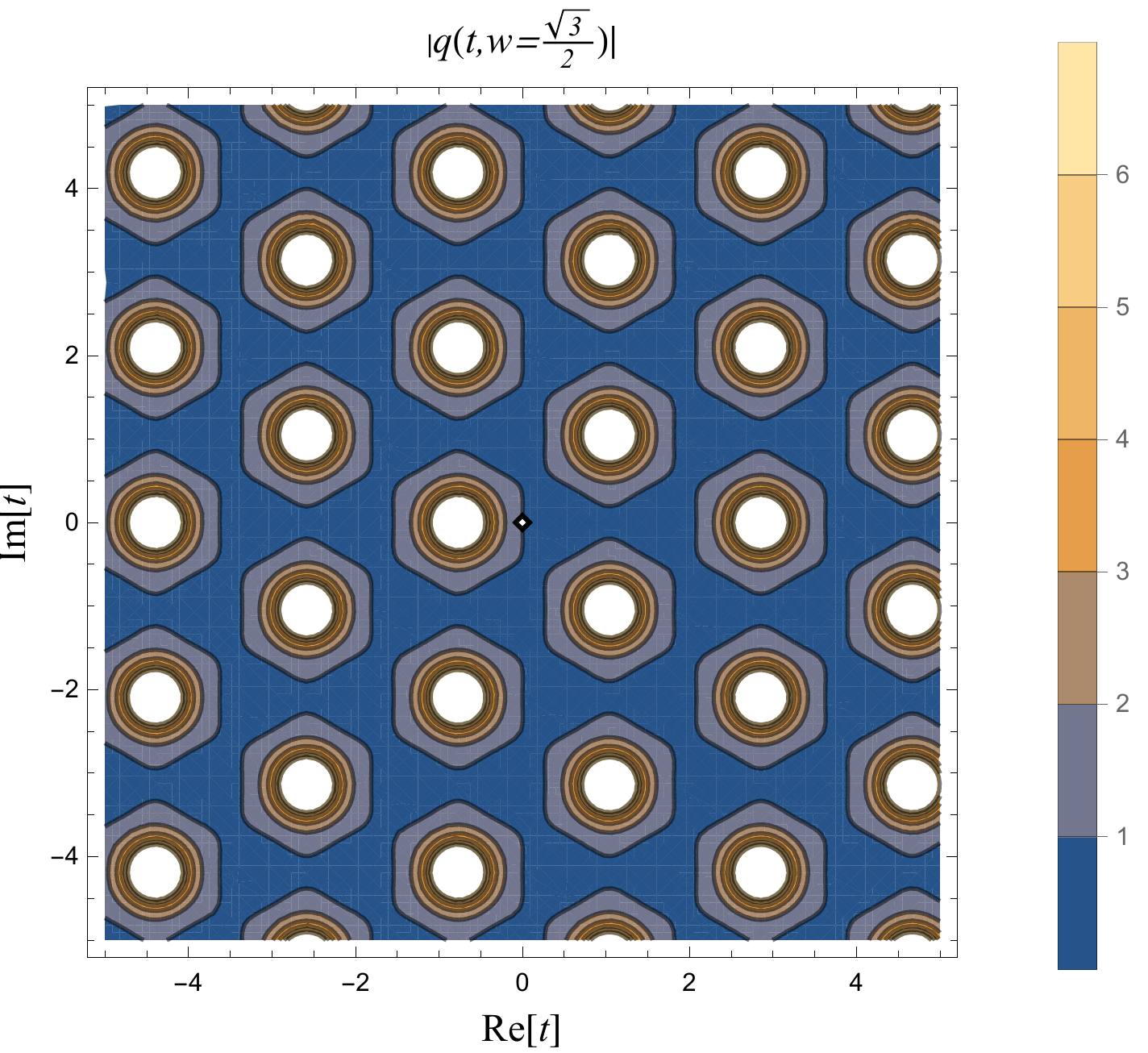}
\caption{$w=\frac{\sqrt{3}}{2}$}
\label{P1}
\end{subfigure}
\hfill
\begin{subfigure}[b]{0.45\textwidth}
\includegraphics[scale=.45]{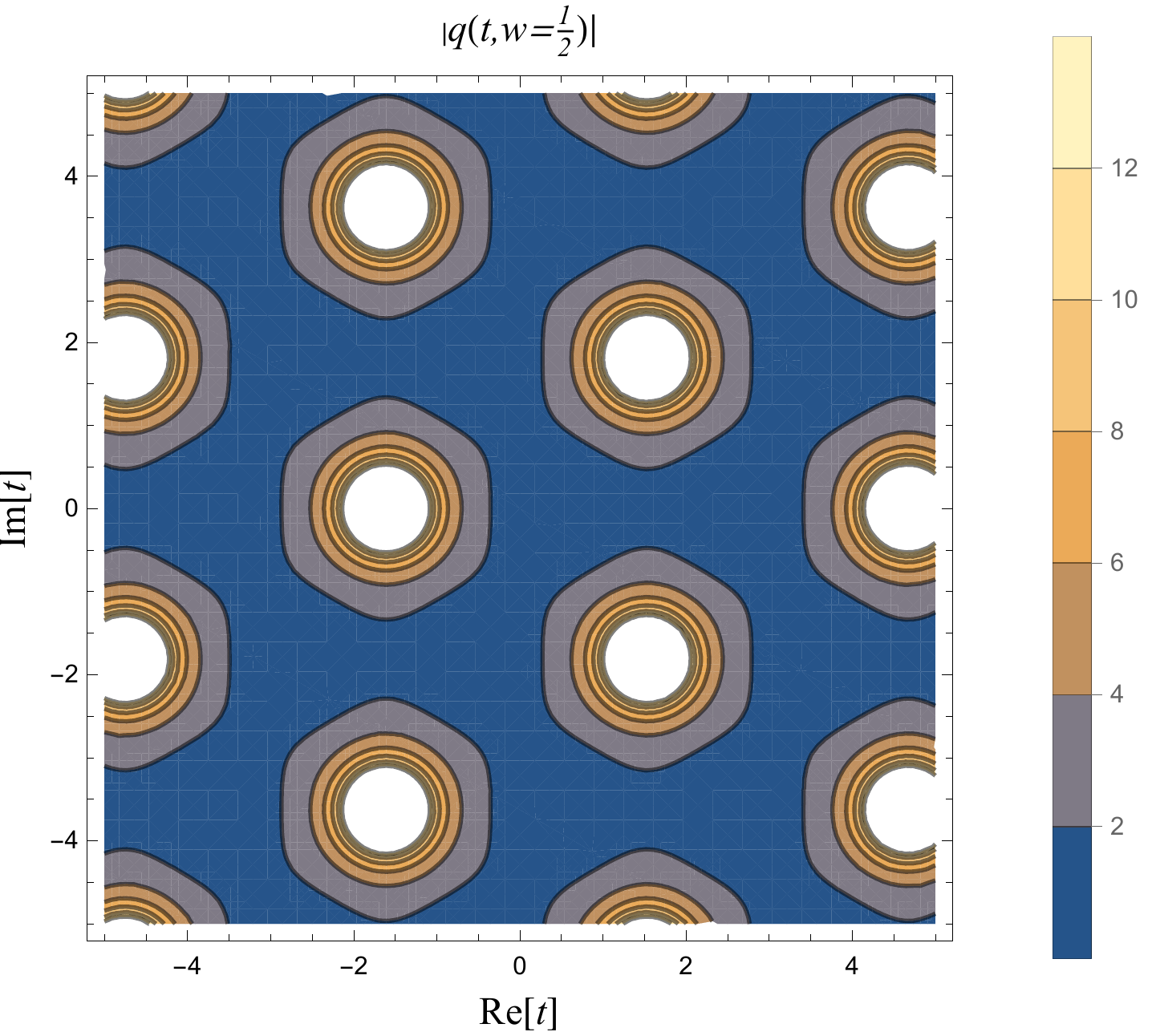}
\caption{ $w=\frac{1}{2}$}
\label{P2}
\end{subfigure}
\hfill
\begin{subfigure}[b]{0.45\textwidth}
\includegraphics[scale=.45]{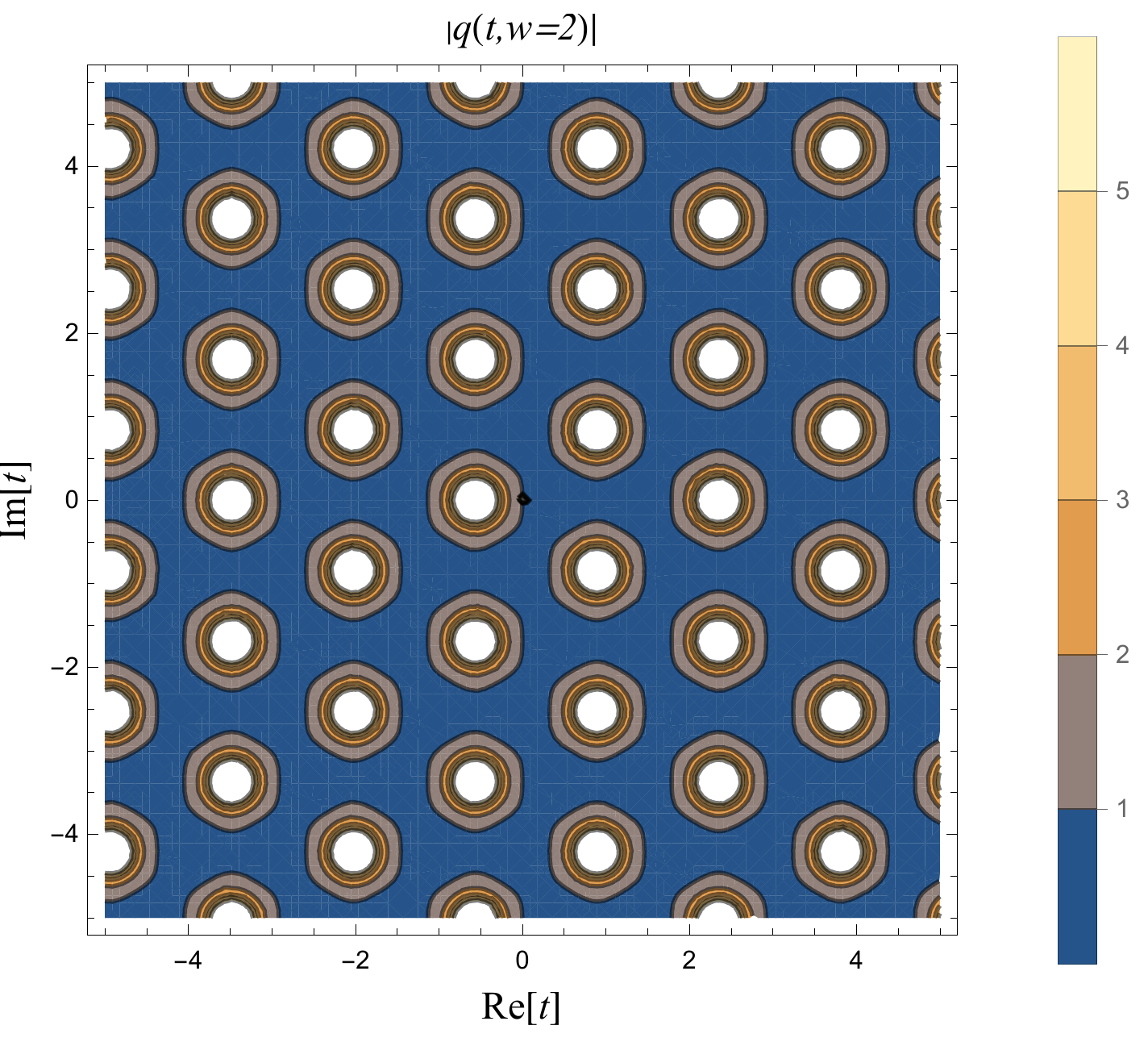}
\caption{$w=2$}
\label{P3}
\end{subfigure}
\hfill
\begin{subfigure}[b]{0.45\textwidth}
\includegraphics[scale=.45]{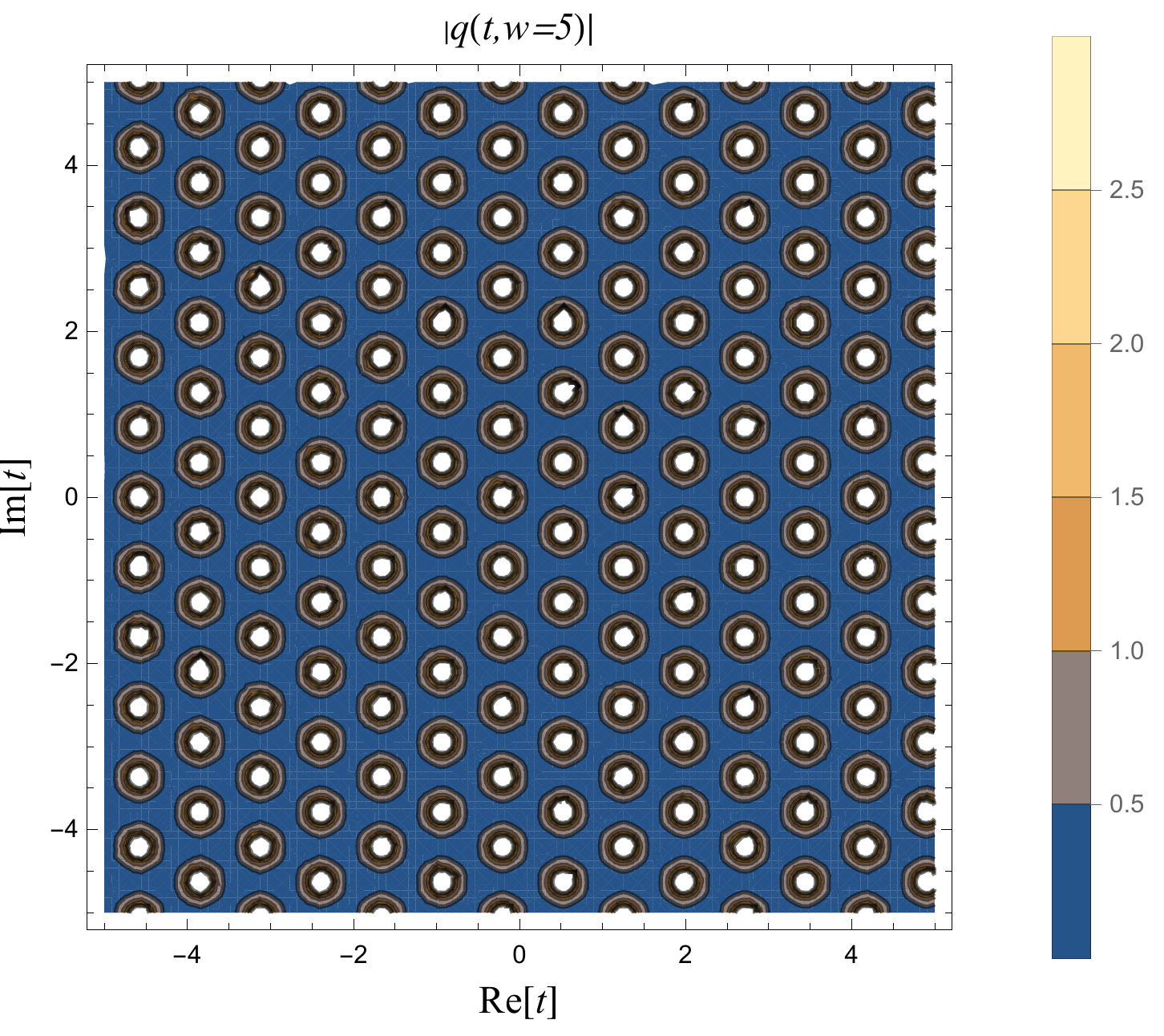}
\caption{$w=5$}
\label{P4}
\end{subfigure}
\caption{The function $|q(t,w)|$ in the case $N=1$ is plotted for different values of $w$. It can be seen that it exhibits a double periodic pole structure, where the distance between the poles is controlled by the variable $w$.}
\label{fig:N1}
\end{figure}

Figure~\ref{fig:N1} shows the function  $q(t,w)$ for the values $w=\frac{\sqrt{3}}{2}$ (\ref{P1}), $w=\frac{1}{2}$ (\ref{P2}),$w=2$ (\ref{P3}) and $w=5$ (\ref{P4}). It can be seen, that the loop function possesses periodically located poles in the complex $t$-plane. The variable $w$ controls the distance between these poles,
the higher values of $|1-w|$ result in the smaller distances between poles.

\section{Conclusions}
In this paper we performed
the all-order summation of
leading logs in the two-dimensional $O(N+1)/O(N)$ $\sigma$-model deformed by the most general dimension-four operators, see Eq.~(\ref{LagrangianBissextile2}).
This theory
includes both renormalizable and non-renormalizable interactions. As a special case, it contains the $T\bar{T}$-deformation that is of particular interest.
For the special case of the $T\bar{T}$-deformation the LL amplitude vanishes for all $N$, validating its integrability property. For the most general deformation
of the theory by the irrelevant dimension-four operators (see Eq.~(\ref{LagrangianBissextile2})) we 
established non-linear differential equations for the LL amplitude
to arbitrary loop order. These equations can be viewed as the generalization of RGE for the case of 2D non-renormalizable field theories.

The corresponding equations allowed us to build a systematic expansion of the LL amplitude around the $T\bar{T}$-deformation of the theory.
In the theory (\ref{LagrangianBissextile2}) the corresponding expansion is governed by the small parameter $1/G$. The developed method allows
to obtain  the all-loop resumed scattering amplitude to a given order in $1/G$, thus providing 
a tool to study the properties of
the theory deformed by the most general dimension-four irrelevant operators.

For several values of the group parameter $N$ the exact LL amplitude is obtained, without the expansion in the perturbation around
the $T\bar{T}$-deformed theory, {\it i.e.} for
arbitrary
value of $1/G$. The obtained exact results for the
$2 \to 2$
LL amplitude provide
new examples of the quasi-renormalizable field theories introduced in \cite{Polyakov:2018rdp}. The $2 \to 2$ LL amplitude in these cases
turns to be a non-trivial meromorphic function of variable $z=\frac{s}{4\pi G}\ln\left(\mu^2/s\right)$ which we found explicitly for the cases $N=-2,0,1,2,\infty$.

Probably the most interesting and physics relevant is the case of $N=2$.
The LL amplitude for $N=2$ possesses an infinite number of equidistant poles in the variable $z$.
Interestingly enough, the location of these poles qualitatively changes when the variable $w=G/(s F^2)$ crosses unity:
for $w< 1$ they are located  equidistantly along the imaginary $z$-axis, while for $w>1$ along the real $z$-axis. The distance between poles
increases as $w\to 1$.  It would be extremely interesting to understand the physical implications of this qualitative change
of the regime for physical systems described by the corresponding non-linear $\sigma$-model.

\renewcommand{\thesection}{}
\makeatletter
\def\@seccntformat#1{\csname #1ignore\expandafter\endcsname\csname the#1\endcsname\quad}

\section{Acknowledgements}
We are grateful to D. Kazakov and A. Vladimirov for the enlightening discussions.
The work of JL and MVP is supported by the BMBF (grant 05P2018). Research of NSS at Perimeter Institute is supported in part by the Government of Canada through the Department of Innovation, Science and Economic Development and by the Province of Ontario through the Ministry of Colleges and Universities.

\appendix
\section{Notes on the $O(N)$-group}
\label{Group}

The fundamental representation of the $O(N)$ group  can be decomposed into three irreducible subspaces according to the projection operators
\begin{align}
\begin{aligned} P_{a b c d}^{I=0} &=\frac{1}{N} \delta_{a b} \delta_{d c} ; \quad P_{a b c d}^{I=1}=\frac{1}{2}\left(\delta_{a d} \delta_{b c}-\delta_{a c} \delta_{b d}\right); \\ P_{a b c d}^{I=2} &=\frac{1}{2}\left(\delta_{a d} \delta_{b c}+\delta_{a c} \delta_{b d}\right)-\frac{1}{N} \delta_{a b} \delta_{c d}; \end{aligned}
\end{align}
satisfying the completeness relation
\begin{align}
P_{a b c d}^{I=0}+P_{a b c d}^{I=1}+P_{a b c d}^{I=2}=\delta_{a d} \delta_{b c}.
\end{align}
The crossing symmetry   mixes
the different isospin channels and can be formulated as
\begin{align}
\begin{aligned} \mathcal{M}^{I}(s, t, u) &=C_{s t}^{I J} \mathcal{M}^{J}(t, s, u); \\ \mathcal{M}^{I}(s, t, u) &=C_{s u}^{I J} \mathcal{M}^{J}(u, t, s); \\ \mathcal{M}^{I}(s, t, u) &=C_{t u}^{I J} \mathcal{M}^{J}(s, u, t), \end{aligned}
\end{align}
where the crossing matrices are defined as
\begin{align}
C_{s u}^{I J}=\frac{1}{d_{I}} P_{a b c d}^{I} P_{b d a c}^{J} ; \quad C_{s t}^{I J}=\frac{1}{d_{I}} P_{a b c d}^{I} P_{c b a d}^{J} ; \quad C_{t u}^{I J}=\frac{1}{d_{I}} P_{a b c d}^{I} P_{b a c d}^{J}
\end{align}
with the corresponding dimensions of invariant subspaces
\begin{align}
d_{I}=P_{a b b a}^{I}=\left\{1, \frac{N(N-1)}{2}, \frac{(N+2)(N-1)}{2}\right\}.
\end{align}
The explicit form of the crossing matrices reads
\be
&&
C_{s u}=\left(\begin{array}{ccc}{\frac{1}{N}} & {\frac{1-N}{2}} & {\frac{N^{2}+N-2}{2 N}} \\ {-\frac{1}{N}} & {\frac{1}{2}} & {\frac{N+2}{2 N}} \\ {\frac{1}{N}} & {\frac{1}{2}} & {\frac{N-2}{2 N}}\end{array}\right);\quad C_{s t}=\left(\begin{array}{ccc}{\frac{1}{N}} & {\frac{N-1}{2}} & {\frac{N^{2}+N-2}{2 N}} \nn \\ {\frac{1}{N}} & {\frac{1}{2}} & {-\frac{N+2}{2 N}} \\ {\frac{1}{N}} & {-\frac{1}{2}} & {\frac{N-2}{2 N}}\end{array}\right);\\ &&
C_{t u}=\left(\begin{array}{ccc}{1} & {0} & {0} \\ {0} & {-1} & {0} \\ {0} & {0} & {1}\end{array}\right).
\ee

\section{Expression of LL scattering amplitudes through the  loop function $\Omega(z,w)$}
\label{OmegavsA}

Here we provide the expressions for the transmission
($T$)
and reflection
($R$)
LL amplitudes for all isospin channels in terms of the loop function
$\Omega(z,w)$
(\ref{eq:omdef}).
\be
{\cal M}^{I=0, \; T}(s)&=&
\frac{s}{F^2}(N-1)-4s^2\lambda+\frac{s^2}{ G} (N+3)+\frac{s^2}{G}\ \Omega\left( \frac{s}{4\pi G} \ln\left(\frac{\mu^2}{s}\right),\frac{G}{s F^2}\right); \nn \\
\label{eq:ampVSom}
{\cal M}^{I=1, \; T}(s)&=&
\frac{s}{F^2}
-4s^2\lambda+\frac{s^2}{ G}
-\frac{s^2}{(N-1)G}\ \Omega\left(-\frac{s}{4\pi G} \ln\left(\frac{\mu^2}{s}\right),
{-}
\frac{G}{s F^2}\right);  \\
{\cal M}^{I=2, \; T}
(s)&=&-\frac{s}{F^2}
-4s^2\lambda+\frac{3s^2}{G} \nn \\
&-&\frac{2s^2}{(N+2)(N-1)G}\
\left[\Omega\left(\frac{s}{4\pi G} \ln\left(\frac{\mu^2}{s}\right),\frac{G}{s F^2}\right)
-\frac{N}{2}\Omega\left( -\frac{s}{4\pi G} \ln\left(\frac{\mu^2}{s}\right),
{-}
\frac{G}{s F^2}\right)\right]; \nn\\
{\cal M}^{I, \; R}(s)&=&(-1)^I\ {\cal M}^{I, \; T}(s).\nn
\ee

\phantomsection

\end{document}